\documentclass{article}

\usepackage{microtype}
\usepackage{graphicx}
\usepackage{subfigure}
\usepackage{booktabs} 
\usepackage{adjustbox}

\usepackage{hyperref}

\usepackage[accepted]{icml2025}

\usepackage{amsmath}
\usepackage{amssymb}
\usepackage{mathtools}
\usepackage{amsthm}
\usepackage{multirow}
\usepackage{calc}

\usepackage[capitalize,noabbrev]{cleveref}

\theoremstyle{plain}

\theoremstyle{definition}

\theoremstyle{remark}

\usepackage[textsize=tiny]{todonotes}

\icmltitlerunning{DeepChem-Variant}

\begin{document}

\twocolumn[
\icmltitle{DeepChem-Variant: \\
A Modular Open Source Framework for Genomic Variant Calling}




\begin{icmlauthorlist}
\icmlauthor{Ankita Vaishnobi Bisoi}{ins,comp}
\icmlauthor{Shreyas V}{ins,comp}
\icmlauthor{Jose Siguenza}{comp}
\icmlauthor{Bharath Ramsundar}{comp}
\end{icmlauthorlist}

\icmlaffiliation{ins}{BITS Pilani Goa Campus, India}
\icmlaffiliation{comp}{Deep Forest Sciences}

\icmlcorrespondingauthor{Ankita Vaishnobi Bisoi}{f20212306@goa.bits-pilani.ac.in}

\icmlkeywords{Genomic Variant Calling, Open Source, Bioinformatics}

\vskip 0.3in
]




\begin{abstract}
Variant calling is a fundamental task in genomic research for detecting genetic variations such as single nucleotide polymorphisms (SNPs) and insertions or deletions (indels). This paper presents an enhancement to DeepChem \cite{Ramsundar-et-al-2019}, a widely used open source drug discovery framework, through the integration of DeepVariant \cite{poplin2018universal}. We introduce DeepChem-Variant, a variant calling pipeline that leverages DeepVariant's convolutional neural network (CNN) architecture to improve variant detection accuracy and reliability. DeepChem-Variant has stages for realignment of sequencing reads, candidate variant detection, and pileup image generation, followed by variant classification using either the original modified Inception V3  model or our novel MobileNetV2  implementation. We performed 3 case studies to validate our approach. Our work also contributes optimized utility functions for genomic data formats, including enhanced DataLoaders for BAM, SAM, and CRAM files, and an optimized FASTALoader. These implementations collectively provide a modular and extensible variant calling framework within DeepChem, enabling tighter integration of DeepChem's drug discovery infrastructure with bioinformatics pipelines for future research.
\end{abstract}

\section{Introduction}
\label{introduction}

Variant calling identifies single nucleotide polymorphisms (SNPs) and insertions/deletions (indels) from sequencing data, foundational for population genetics, disease etiology, and precision medicine applications including risk prediction and therapeutic interventions. Standard approaches like GATK \cite{mckenna2010genome} and SAMtools \cite{li2009sequence} use probabilistic models that struggle with ambiguous or low-quality data. These methods face challenges in noisy genomic regions, reducing sensitivity and specificity in low-coverage areas or regions with complex structural variations.

DeepVariant \cite{poplin2018universal, poplin2017practical}, developed by Google, uses a Convolutional Neural Network (CNN) \cite{krizhevsky2012imagenet} to reframe the variant calling task as an image classification problem. Pileup images (Section~\ref{sec:2.2}) are generated from sequencing reads and analyzed by the CNN to distinguish true variants from sequencing errors. This approach outperforms traditional heuristic-based methods, achieving higher accuracy in variant detection, particularly across diverse sequencing platforms and in regions where conventional tools exhibit reduced performance.

However, while the code for DeepVariant is accessible on platforms such as GitHub, part of the code is written in C++ and is challenging to modify or extend. The architecture and components are fixed within the provided framework, making it difficult for researchers to adapt DeepVariant to explore novel hypotheses or improve specific sub-components for their experimental needs.

To address the need for modular open-source implementations of computational genomic tools, we integrate an implementation of DeepVariant  as DeepChem-Variant into the DeepChem \cite{Ramsundar-et-al-2019} framework. DeepChem, an open-source Python library designed for scientific machine learning and deep learning, has established itself as a versatile platform for applications in molecular machine learning ranging from the MoleculeNet benchmark suite \cite{wu2018moleculenet} to protein-ligand interaction modeling \cite{gomes2017atomic}, and generative modeling of molecules \cite{frey2022fastflows}, among others. 

Deepchem's modular architecture, comprising components such as data loaders, featurizers, splitters, models, and metrics, provides an extensible system that supports the development of custom workflows. The DeepChem community of developers and contributors actively maintains all implementations in this system, which are validated through continuous integration and delivery (CI/CD) pipelines. The incorporation of DeepChem-Variant into DeepChem significantly broadens its functionality, enabling variant calling workflows to be conducted entirely within an open-source Python ecosystem. We anticipate this infrastructure will enable subsequent computational work exploring the intersection of drug discovery and bioinformatics. 


\begin{figure}[H]
    \centering
    \includegraphics[width=200 pt]{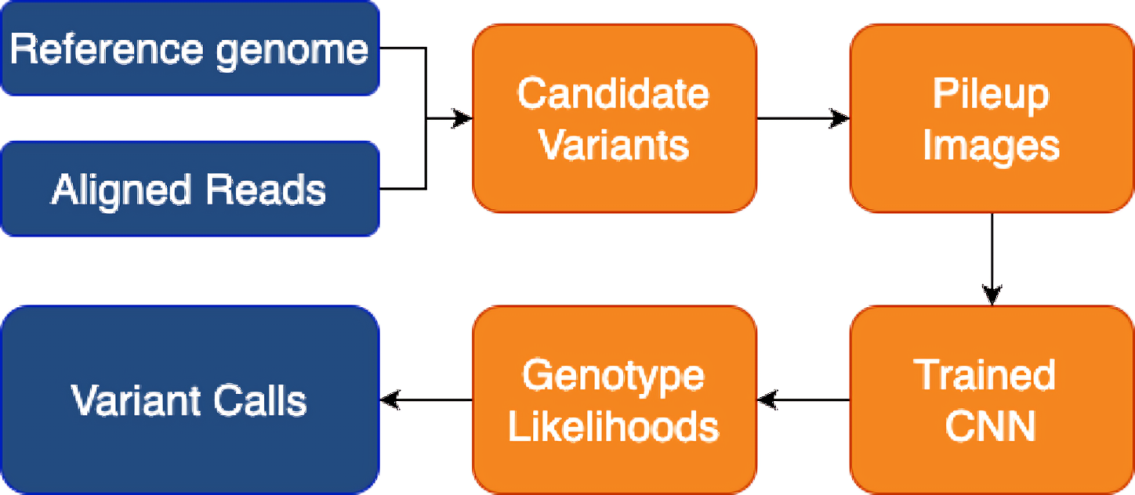} 
    \caption{DeepVariant workflow: Reference genome and aligned reads generate candidate variants, which are converted to pileup images, processed by a trained CNN to produce genotype likelihoods, and finally output as variant calls.}
    \label{fig:dv_pipeline_inceptionv3}
\end{figure}

\section{Implementations}

\begin{figure}[H]
    \centering
    \includegraphics[width=160 pt]{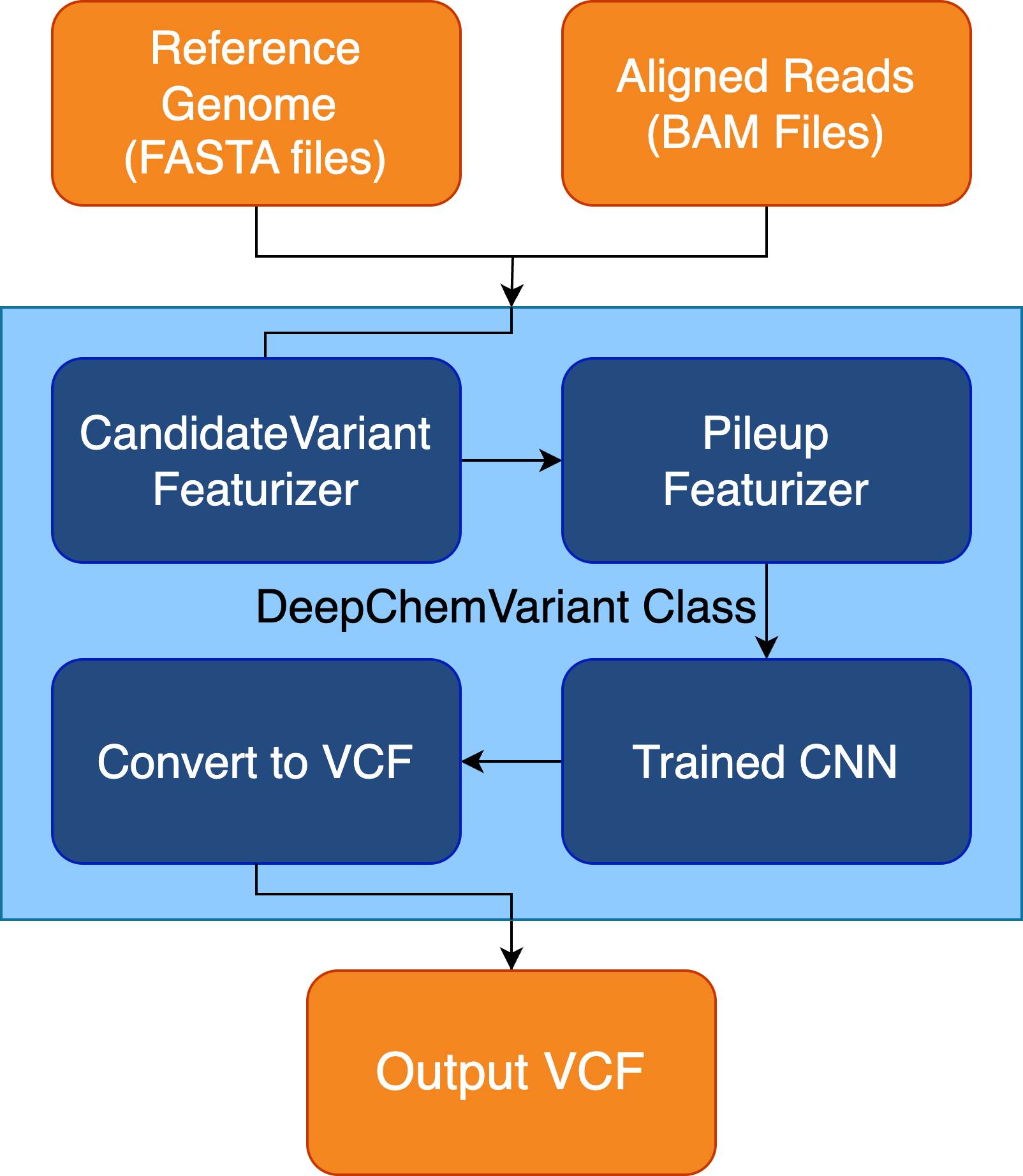}
    \caption{DeepChem-Variant workflow: Reference genome (FASTA) and aligned reads (BAM) feed into the DeepVariant class containing candidate variant featurizer, pileup featurizer, trained CNN (MobileNetV2 or InceptionV3), and VCF converter to produce output Variant Call Format (VCF) files.}
    \label{fig:deepvariant_pipeline_in_deepchem}
\end{figure}

DeepChem-Variant has three primary components: candidate variant detection, pileup image generation, and a deep learning model designed for variant calling. These components are implemented through modular featurizers and a custom convolutional neural network (CNN). To efficiently handle various sequence alignment formats, we developed specialized utility classes including \texttt{BAMLoader}, \texttt{SAMLoader}, and \texttt{CRAMLoader}, enabling seamless integration with diverse genomic datasets. We also enhanced the \texttt{FASTALoader} for efficient reference genome access and significantly optimized the \texttt{FASTAFeaturizer} to process raw nucleotide sequences directly, rather than converting to one-hot encoded representations, resulting in a 210-fold acceleration in processing speed.

\subsection{Candidate Variant Detection}
The first stage involves realigning input sequencing reads and identifying candidate variants through the \texttt{CandidateFeaturizer} class. Reads are provided in BAM\cite{li2009sequence} format, storing compressed alignments of sequencing reads to a reference genome. The featurizer supports optional realignment for improved variant detection accuracy, optional multiprocessing (achieving 8× speedup), and optional labeling for training purposes when VCF ground truth is provided.

The realignment process introduces haplotype awareness through realignment using the \texttt{realign\_read} method, using Striped Smith Waterman algorithm \cite{Zhao2013} to improve read positioning. The \texttt{count\_alleles} method then tallies base occurrences at each position from aligned reads, accounting for CIGAR operations (which encode how sequencing reads align to the reference genome, indicating matches, insertions, deletions, and other alignment operations) to handle insertions, deletions, and matches. The \texttt{detect\_candidates} method identifies potential variants by comparing observed bases against the reference, flagging positions where alternative alleles exceed the minimum count and frequency thresholds. Finally, the \texttt{left\_align\_indel} method standardizes indel representations by shifting them to their leftmost valid positions, ensuring consistent variant notation across samples.

An optimized Smith-Waterman alignment algorithm performs realignment using PyTorch's GPU-accelerated tensor operations. Unlike traditional optimized implementations in C/C++, this approach exploits Python's high-level interface while maintaining competitive performance through PyTorch's CUDA backend. Alignment scores are computed using vectorized operations with substitution scores derived from binary match/mismatch masks.

Left-alignment of indels proves critical for variant standardization. Without left-alignment, candidate detection yielded inconsistent results across samples (average: 342,536 variants), while left-alignment produced 1,727,087 average standardized candidates, demonstrating the importance of variant normalization.

The \texttt{min\_count} parameter, which is the minimum number of reads that must support an alternate allele for it to be considered a candidate variant, significantly impacts candidate sensitivity and computational efficiency. It helps filter out sequencing errors and reduces false positives by requiring multiple independent observations of the same change. 

\begin{table}[!t]
\centering
\caption{Impact of min\_count parameter on candidate detection using HG004 (GIAB consortium data, using Novaseq whole exome sequencing with IDT capture at 100x coverage)\label{tab:min_count_impact}}%
\resizebox{\columnwidth}{!}{%
\begin{tabular}{@{}c c p{4cm}@{}}
\toprule
min\_count & Total Candidates & Computational Impact \\
\midrule
1 & 8,420,830 & High downstream execution time \\
2 (default) & 1,770,108 & Optimal balance \\
\bottomrule
\end{tabular}%
}
\end{table}

The default \texttt{min\_count=2} provides optimal balance, as reducing to 1 substantially increases downstream processing without meaningful recall improvement. This stage is implemented using the pysam \cite{Gilman_Janzou_Guittet_Freeman_DiOrio_Blair_Boyd_Neises_Wagner_2019} library for efficient BAM and FASTA file manipulation.

\begin{figure}[ht!]
    \centering
    \includegraphics[width=180 pt]{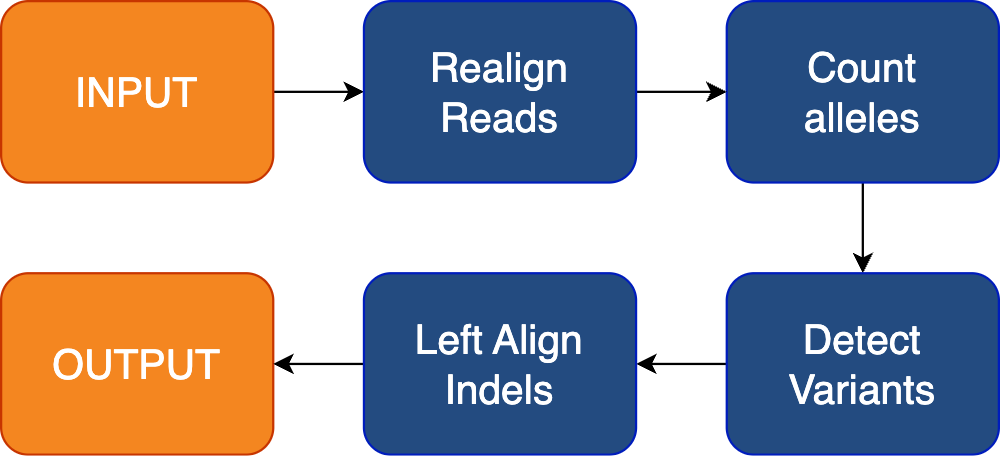}
    \caption{\texttt{CandidateFeaturizer} workflow: Input data flows through realignment of reads, allele counting at genomic positions, variant detection based on frequency thresholds, and left-alignment of indels to produce standardized candidate variants as output.}
    \label{fig:realigner_featurizer}
\end{figure}
\vspace{-0.5cm}

\subsection{Pileup Image Generation}
\label{sec:2.2}
Once candidate variants have been identified, the next stage involves generating pileup images, a core feature of DeepVariant. A pileup image represents aligned sequencing reads at a genomic region centered on a candidate variant position. Each row corresponds to a read, and each column represents a genomic coordinate within the pileup window. 

The \texttt{PileupFeaturizer} is responsible for creating these images, with six channels encoding different features of the sequencing data. Channel 0 encodes base intensities, while Channel 1 captures base quality information. Channel 2 encodes mapping quality, and Channel 3 represents the strand orientation (i.e., whether the read is from the forward or reverse strand). Channel 4 indicates whether the read supports a variant, and Channel 5 encodes the difference between the read and the reference sequence. These multi-channel images provide a rich representation of the underlying sequencing data, enhancing the ability of the deep learning model to distinguish between true variants and sequencing artifacts.

\subsection{Modified CNN for Variant Calling}
The deep learning model employed in this workflow is a custom CNN, derived from either the  \cite{szegedy2015rethinkinginceptionarchitecturecomputer} architecture or MobileNetV2 \cite{sandler2018mobilenetv2}, which is specifically tailored for genomic variant calling from pileup images. We integrated both Inception V3 and MobileNetV2 architectures into DeepChem's core model library. The Inception V3 model's convolutional layers are modified to handle the six-channel input format, while our MobileNetV2 implementation leverages its efficient inverted residual structure and linear bottlenecks for improved computational efficiency without sacrificing accuracy. Both models output a probability score for each candidate variant, indicating whether it is a true variant or a sequencing error. The models integrate into DeepChem's model library, allowing users to easily swap between Inception V3 and MobileNetV2 implementations if desired or integrate variant calling into larger machine learning workflows, such as multi-task learning frameworks or pipelines incorporating other types of genomic data.

\begin{figure}[H]
    \centering
    \includegraphics[width=200 pt]{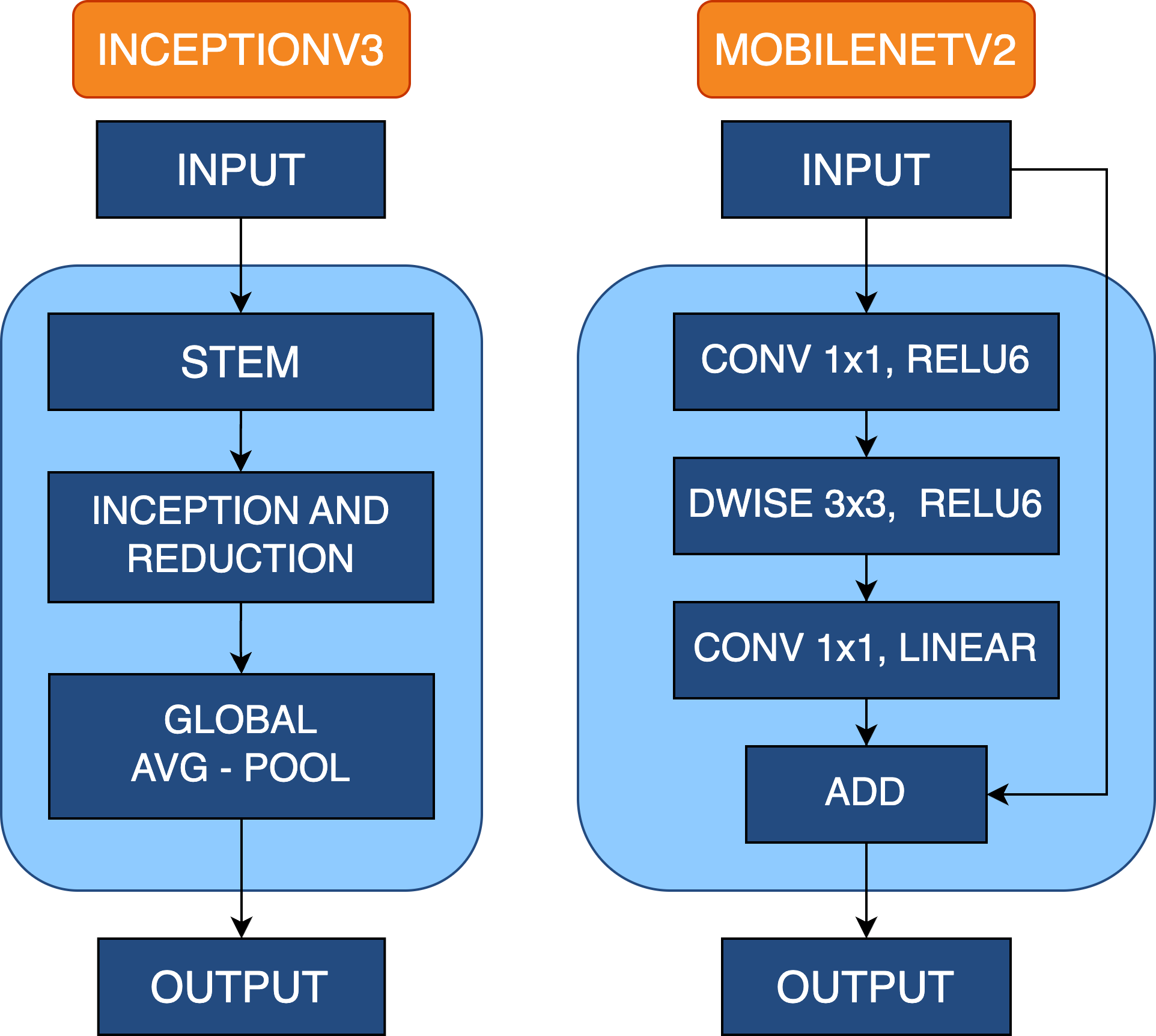} 
    \caption{Architectures of Inception V3 and MobileNetV2 as used in DeepVariant.}
    \label{fig:dv_inceptionv3}
\end{figure}
\vspace{-0.8cm}

\section{Case Studies}

We validated DeepChem-Variant across three genomic contexts using DeepVariant as baseline, since no established ground truth exists for these specialized datasets. DeepChem-Variant values in Table~\ref{tab:combined_comparison} represent the subset of DeepVariant calls that our method successfully detected. 
Sensitivity measures DeepChem-Variant's ability to recover DeepVariant's calls, calculated as the percentage of DeepVariant variants successfully identified by DeepChem-Variant.
VCF outputs from both methods were compared by intersecting variant positions and alleles between them.

\textbf{CRISPR Off-Target Detection:}
Synthetic datasets simulated CRISPR-Cas9 \cite{jinek2012programmable} off-target effects at PAM sites (protospacer adjacent motifs) with insertion, deletion, and random edit patterns across NNNNGATT and NGG motifs. Off-target detection is critical for CRISPR therapeutic safety, as unintended edits can cause harmful mutations.

\textbf{Ancient DNA Analysis:}  
Simulated characteristic ancient DNA damage patterns including C→T transitions, fragmentation, and age-dependent preservation effects spanning 100 to 50,000 years. Ancient DNA analysis enables evolutionary studies and population history reconstruction but requires specialized variant calling due to extreme degradation.

\textbf{Population Genomics:}
Analyzed whole exome sequencing from adult female, adult male, and pediatric male samples to assess demographic-specific variant detection. Population genomics applications require consistent variant calling across diverse samples for disease association studies and personalized medicine.

\vspace{-0.5cm}
\begin{table}[H]
\centering
\caption{Variant detection sensitivity across genomic contexts\label{tab:combined_comparison}}
\resizebox{\columnwidth}{!}
{
\begin{tabular}{@{}l l c c c@{}}
\toprule
Context & Sample/Type & DeepVariant & DeepChem-Variant & Sensitivity (\%) \\
\midrule
\multirow{6}{*}{CRISPR} & NNNNGATT Insertion & 2103 & 1717 & 81.6 \\
& NNNNGATT Deletion & 2284 & 1819 & 79.6 \\
& NNNNGATT Random & 2051 & 1623 & 79.1 \\
& NGG Insertion & 284 & 251 & 88.4 \\
& NGG Deletion & 312 & 282 & 90.4 \\
& NGG Random & 298 & 274 & 91.9 \\
\midrule
\multirow{3}{*}{Ancient DNA} & Recent (100y) & 32,847 & 30,942 & 94.2 \\
& Medieval (800y) & 51,293 & 47,251 & 92.1 \\
& Neanderthal (50,000y) & 78,164 & 68,503 & 87.6 \\
\midrule
\multirow{3}{*}{Population} & Adult Female & 61,245 & 55,732 & 91.0 \\
& Adult Male & 59,874 & 54,605 & 91.2 \\
& Pediatric Male & 60,298 & 54,992 & 91.2 \\
\bottomrule
\end{tabular}
}
\vspace{-0.45cm}
\end{table}

\section{Discussion}

DeepChem-Variant marks a progression in the development of open-source tools for genomic variant calling. By embedding advanced deep learning techniques within a flexible machine learning framework, this integration improves accessibility and customizability of variant calling infrastructure for a wide range of research applications. DeepChem's modular architecture enables easy adaptation, allowing researchers to explore new methodologies, within genomic data analysis workflows.

\subsection{Performance and efficiency}
\label{sec:disc:perf_eff}
The original DeepVariant combined C++ and Python, creating complexity requiring proficiency in both languages. DeepChem-Variant's all-Python implementation (approximately 1,500 lines versus DeepVariant's 35,000 as mentioned in Table~\ref{tab:loc_comparison}) simplifies development, reduces barriers to entry, and leverages Python's scientific computing ecosystem for easier extensibility and rapid prototyping.
DeepChem-Variant offers two CNN architectures: Inception V3 
and MobileNetV2 
Our models were trained on HG001, HG002, HG004, and HG005 WES deduplicated samples at 100x coverage from IDT NovaSeq (300 million rows), compared to production DeepVariant models trained on 8-9 fold larger multi-technology GIAB datasets (2.6 billion rows) \cite{covering_all_bases} and all models were validated on HG003.
The code was implemented in PyTorch \cite{paszke2019pytorch} on Google Colab \cite{googlecolab}. More details about hyperparameters and system specifications are mentioned in Appendix \ref{app:expinfo}, details about datasets are mentioned in Appendix \ref{app:datasets}.  


\begin{table}[!t]
\centering
\caption{Comparison of lines of code between DeepVariant and DeepVariant (DeepChem)\label{tab:loc_comparison}}%
\resizebox{\columnwidth}{!}{%
\begin{tabular}{@{}l p{3 cm} r@{}}
\toprule
Method & Language(s) & Lines of Code (approx.) \\
\midrule
DeepVariant & C++/Python & ~35,000 \\
DeepChem-Variant  & Python & ~1,500 \\
\bottomrule
\end{tabular}%
}
\vspace{-0.55cm}
\end{table}

\subsection{Limitations and future work}
An evaluation of performance metrics (Table~\ref{tab:performance_metrics}) indicates some discrepancies between the original DeepVariant implementation and DeepChem-Variant. 
This is due to limited training data diversity in our experiments compared to production DeepVariant (discussed in section \ref{sec:disc:perf_eff}).
The MobileNetV2 (3.4 million parameters) results are particularly notable given its significantly fewer parameters compared to InceptionV3 (24 million parameters) and lower ImageNet accuracy, yet achieving competitive performance in genomic variant calling. Despite these limitations, our modular open-source Python implementation allows users to easily swap components, such as the CNN architecture or realignment algorithm, as new methods and technologies emerge.

\begin{table}[!t]
\centering
\caption{Performance comparison of variant calling methods\label{tab:performance_metrics}}%
\resizebox{\columnwidth}{!}{%
\begin{tabular}{@{}l l c c c@{}}
\toprule
Method & Variant Type & Recall & Precision & F1-Score \\
\midrule
DeepVariant & INDEL & 0.971 & 0.993 & 0.982 \\
DeepVariant & SNP & 0.988 & 0.998 & 0.993 \\
\midrule
DeepChem-Variant (MobileNetV2) & INDEL & 0.912 & 0.934 & 0.923 \\
DeepChem-Variant (MobileNetV2) & SNP & 0.922 & 0.941 & 0.943 \\
\midrule
DeepChem-Variant (InceptionV3) & INDEL & 0.923 & 0.951 & 0.933 \\
DeepChem-Variant (InceptionV3) & SNP & 0.931 & 0.954 & 0.939 \\
\bottomrule
\end{tabular}%
}
\vspace{-0.5cm}
\end{table}

\section{Conclusion}
\label{sec:conclusion}

In this work, we introduce DeepChem-Variant which enables researchers to utilize advanced deep learning methods for genomics within a customizable Python framework, expanding machine learning applications in genomics. 
While performance differences compared to the original CNN implementation were observed, due to training on smaller datasets and architectural choices, the integration within DeepChem facilitates rapid future improvements. This will also allow for the easier incorporation of novel genomic analysis methodologies. We note that the observed performance reflects the current training data and model architecture, and leave further optimization for future work. As an open-source tool, we anticipate community contributions will drive further enhancements, ultimately benefiting areas such as personalized medicine and population genetics.

\section*{Impact Statement}

This paper presents work whose goal is to advance the field of Machine Learning. There are many potential societal consequences of our work, none which we feel must be specifically highlighted here.

\bibliography{example_paper}
\bibliographystyle{icml2025}

\newpage
\appendix
\onecolumn

\section{Implementation Details}

This project added significant functionality to DeepChem through new classes and enhancements spanning featurizers, data loaders, models, and variant calling infrastructure. Table~\ref{tab:code_contributions} summarizes the key contributions.

\begin{table} [H]
\centering
\caption{Code contributions to DeepChem framework\label{tab:code_contributions}}%
\resizebox{\columnwidth}{!}{%
\begin{tabular}{@{}l l p{10cm}@{}}
\toprule
Class Name & Parent Class & Description \\
\midrule
\texttt{SAMFeaturizer} & \texttt{Featurizer} & Processes SAM alignment files \\
\texttt{BAMFeaturizer} & \texttt{Featurizer} & Processes BAM alignment files \\
\texttt{CRAMFeaturizer} & \texttt{Featurizer} & Processes CRAM alignment files \\
\texttt{FASTAFeaturizer} & \texttt{Featurizer} & Enhanced sequence processing with 210× speedup \\
\midrule
\texttt{SAMLoader} & \texttt{DataLoader} & Loads SAM format files \\
\texttt{BAMLoader} & \texttt{DataLoader} & Loads BAM format files \\
\texttt{CRAMLoader} & \texttt{DataLoader} & Loads CRAM format files \\
\texttt{FASTALoader} & \texttt{DataLoader} & Loads reference genome sequences \\
\midrule
\texttt{MobileNetV2Model} & \texttt{TorchModel} & Efficient CNN for variant classification \\
\texttt{InceptionV3Model} & \texttt{TorchModel} & High-accuracy CNN for variant classification \\
\midrule
\texttt{CandidateFeaturizer} & \texttt{Featurizer} & Identifies potential variant sites \\
\texttt{PileupFeaturizer} & \texttt{Featurizer} & Generates multi-channel alignment images \\
\texttt{DeepChemVariant} & \texttt{TorchModel} & Complete variant calling pipeline \\
\bottomrule
\end{tabular}%
}
\end{table}

\section{Hyperparameters and System Specifications}
\label{app:expinfo}
 DeepChem-Variant (both InceptionV3 and MobileNetV2) utilized a pileup image representation with a window size of 221 base pairs, capturing 100 read depths across 6 channels encoding base identity, base quality, mapping quality, strand orientation, variant support, and reference match indicators. Training was conducted using the Adam optimizer with a learning rate of 1e-3, batch size of 128, and 10 epochs. The model was trained on Google Colab's L4 GPU infrastructure with 8 data loader workers for parallel data processing. Variant candidates were filtered using minimum allele count and frequency thresholds of 2 reads and 1\% respectively, ensuring adequate support for downstream classification while maintaining computational efficiency.

\section{Datasets}
\label{app:datasets}

DeepChem-Variant (both InceptionV3 and MobileNetV2) was trained on a constrained dataset comprising HG001, HG002, HG004, and HG005 WES deduplicated samples at 100x coverage from a single sequencing platform (IDT NovaSeq, 12 GB total). In contrast, production DeepVariant models utilize extensive multi-technology GIAB datasets that are ~8-9 fold larger, incorporating data from diverse sequencing platforms including HiSeqX, NovaSeq, and PCR-positive samples across multiple WES capture kits, different sequencing depths, and various sample preparation methods \cite{covering_all_bases}. 
This substantial difference in training data scale, technological diversity, and sample heterogeneity results in reduced performance due to our model's limited exposure to the full spectrum of sequencing artifacts and variant patterns present in real-world genomic data.

\section{Computation of Candidate Variants}
\subsection{CandidateVariantFeaturizer}
The CandidateVariantFeaturizer algorithm processes genomic data in sliding windows to identify potential variant sites. For each genomic region, it extracts aligned reads and reference sequences, optionally performs Smith-Waterman realignment to improve accuracy, then counts allele frequencies at each position. Candidate variants are detected by comparing observed alleles against the reference using minimum count and frequency thresholds. Finally, indel variants are left-aligned to ensure standardized representation. The algorithm returns an array of candidate variants with associated metadata for downstream analysis.

\begin{algorithm}
\caption{CandidateVariantFeaturizer}
\label{alg:candidatefeaturizer}
\begin{algorithmic}
\STATE \textbf{Input:} BAM file $B$, FASTA file $F$
\STATE \textbf{Output:} Candidate variant array
\FOR{each region $(chrom, start, end)$}
    \STATE $reads \gets \texttt{fetch}(B, chrom, start, end)$ /* extract aligned reads */
    \STATE $ref\_seq \gets \texttt{fetch}(F, chrom, start, end)$ /* extract reference sequence */
    \IF{realign enabled}
        \STATE $reads \gets \texttt{smith\_waterman\_realign}(reads, ref\_seq)$ /* improve alignment accuracy */
    \ENDIF
    \STATE $counts \gets \texttt{count\_alleles}(reads, ref\_seq)$ /* tally base frequencies */
    \STATE $variants \gets \texttt{detect\_candidates}(counts, \text{min\_count}, \text{min\_frac})$ /* identify variant sites */
    \FOR{each variant $v \in variants$}
        \STATE $v \gets \texttt{left\_align\_indel}(v)$ /* standardize representation */
        \STATE \texttt{output.append}(v) /* add to result set */
    \ENDFOR
\ENDFOR
\STATE \textbf{Return:} candidate variants with metadata
\end{algorithmic}
\end{algorithm}

\subsection{Realignment of Reads}

\begin{algorithm}
\caption{Smith-Waterman Alignment}
\label{alg:smithwaterman}
\begin{algorithmic}
\STATE \textbf{Input:} Query sequence $Q$, reference sequence $R$
\STATE \textbf{Output:} Aligned query sequence
\STATE $H, E, F \gets \texttt{zeros}(|Q|+1, |R|+1)$ /* alignment, gap matrices */
\STATE $pointer \gets \texttt{zeros}(|Q|+1, |R|+1)$ /* traceback directions */
\STATE $max\_score, max\_pos \gets 0, (0,0)$ /* track optimal alignment */
\FOR{$i = 1$ to $|Q|$}
    \FOR{$j = 1$ to $|R|$}
        \STATE $match \gets H[i-1,j-1] + \texttt{score}(Q[i], R[j])$ /* diagonal score */
        \STATE $E[i,j] \gets \max(H[i-1,j] + gap\_open, E[i-1,j] + gap\_extend)$ /* vertical gap */
        \STATE $F[i,j] \gets \max(H[i,j-1] + gap\_open, F[i,j-1] + gap\_extend)$ /* horizontal gap */
        \STATE $H[i,j] \gets \max(0, match, E[i,j], F[i,j])$ /* local alignment score */
        \IF{$H[i,j] > max\_score$}
            \STATE $max\_score, max\_pos \gets H[i,j], (i,j)$ /* update maximum */
        \ENDIF
    \ENDFOR
\ENDFOR
\STATE $aligned \gets \texttt{traceback}(pointer, max\_pos, Q)$ /* reconstruct alignment */
\STATE \textbf{Return:} $aligned$
\end{algorithmic}
\end{algorithm}
The Smith-Waterman algorithm performs optimal local sequence alignment using dynamic programming. It initializes three scoring matrices: H for alignment scores, E and F for gap penalties. The algorithm fills these matrices by calculating match/mismatch scores and gap costs, maintaining traceback pointers for reconstruction. It identifies the maximum local alignment score during matrix filling, then traces back from this position to reconstruct the optimal alignment path. This implementation uses PyTorch tensors for vectorized operations, providing GPU acceleration while maintaining the algorithm's quadratic time complexity.

\subsection{Counting Alleles}
The \texttt{count\_alleles} funtion tallies base frequencies at each genomic position by processing aligned reads. It initializes a dictionary array to store counts per position, then iterates through each read's CIGAR string to handle matches, insertions, and deletions appropriately. For matched regions, it extracts base calls and increments corresponding position counters. The algorithm advances position pointers based on CIGAR operations, ensuring proper coordinate mapping between read sequences and reference positions.

\begin{algorithm}
\caption{Count Alleles}
\label{alg:countalleles}
\begin{algorithmic}
\STATE \textbf{Input:} Reads $R$, reference sequence $ref$, region start $start$
\STATE \textbf{Output:} Allele counts per position
\STATE $counts \gets \texttt{empty\_dict\_array}(|ref|)$ /* one dict per position */
\FOR{each read $r \in R$}
    \IF{$r$ is unmapped or duplicate}
        \STATE \textbf{continue} /* skip low-quality reads */
    \ENDIF
    \STATE $ref\_pos, query\_pos \gets r.start, 0$ /* initialize positions */
    \FOR{each $(operation, length)$ in $r.cigar$}
        \IF{$operation == \texttt{MATCH}$}
            \FOR{$i = 0$ to $length-1$}
                \STATE $pos \gets ref\_pos + i - start$ /* convert to region coordinates */
                \IF{$0 \leq pos < |ref|$ and $query\_pos + i < |r.sequence|$}
                    \STATE $base \gets r.sequence[query\_pos + i]$ /* extract base call */
                    \STATE $counts[pos][base] \gets counts[pos][base] + 1$ /* increment count */
                \ENDIF
            \ENDFOR
            \STATE $ref\_pos, query\_pos \gets ref\_pos + length, query\_pos + length$ /* advance both */
        \ELSIF{$operation == \texttt{INSERTION}$}
            \STATE $query\_pos \gets query\_pos + length$ /* advance query only */
        \ELSIF{$operation == \texttt{DELETION}$}
            \STATE $ref\_pos \gets ref\_pos + length$ /* advance reference only */
        \ENDIF
    \ENDFOR
\ENDFOR
\STATE \textbf{Return:} $counts$
\end{algorithmic}
\end{algorithm}

\subsection{Detecting candidates}

The \texttt{detect\_candidates} function identifies potential variant sites by applying frequency-based filtering to allele counts. It examines each genomic position with coverage, calculates total read depth, and compares observed alleles against the reference. Variants are flagged as candidates if they differ from the reference allele and exceed both minimum count and frequency thresholds, helping reduce false positives from sequencing errors while retaining true biological variants.

\begin{algorithm}
\caption{Detect Candidate Variants}
\label{alg:detectcandidates}
\begin{algorithmic}
\STATE \textbf{Input:} Allele counts $counts$, reference $ref$, thresholds $min\_count$, $min\_frac$
\STATE \textbf{Output:} Candidate variant list
\STATE $candidates \gets \texttt{empty\_list}()$ /* initialize result list */
\FOR{$i = 0$ to $|counts|-1$}
    \STATE $total \gets \sum(counts[i].values())$ /* sum all allele counts */
    \IF{$total == 0$}
        \STATE \textbf{continue} /* skip positions with no coverage */
    \ENDIF
    \STATE $ref\_base \gets ref[i]$ /* get reference allele */
    \FOR{each $(base, count)$ in $counts[i]$}
        \IF{$base \neq ref\_base$ and $count \geq min\_count$ and $\frac{count}{total} \geq min\_frac$}
            \STATE $candidate \gets (i, ref\_base, base, count, total)$ /* create variant record */
            \STATE $candidates.\texttt{append}(candidate)$ /* add to candidates */
        \ENDIF
    \ENDFOR
\ENDFOR
\STATE \textbf{Return:} $candidates$
\end{algorithmic}
\end{algorithm}

\subsection{Left Aligning Indels}

The left align indels algorithm standardizes indel representation by shifting variants to their leftmost valid position. It first checks if the variant is a SNP (equal lengths or different first bases), returning unchanged if so. For indels, it trims common prefix and suffix sequences, then iteratively shifts the variant leftward by comparing flanking bases from the reference genome until no further movement is possible, ensuring consistent variant notation across different calling methods.

\begin{algorithm}
\caption{Left Align Indels}
\label{alg:leftalign}
\begin{algorithmic}
\STATE \textbf{Input:} Chromosome $chrom$, position $pos$, reference allele $ref$, alternate allele $alt$, FASTA $fasta$
\STATE \textbf{Output:} Left-aligned position and alleles
\IF{$|ref| == |alt|$ or $ref[0] \neq alt[0]$}
    \STATE \textbf{Return:} $(pos, ref, alt)$ /* SNP, no alignment needed */
\ENDIF
\STATE $seq, seq\_alt, left \gets ref, alt, pos$ /* initialize working variables */
\WHILE{$|seq| > 1$ and $|seq\_alt| > 1$ and $seq[-1] == seq\_alt[-1]$}
    \STATE $seq, seq\_alt \gets seq[:-1], seq\_alt[:-1]$ /* trim common suffix */
\ENDWHILE
\WHILE{$|seq| > 1$ and $|seq\_alt| > 1$ and $seq[0] == seq\_alt[0]$}
    \STATE $seq, seq\_alt \gets seq[1:], seq\_alt[1:]$ /* trim common prefix */
    \STATE $left \gets left + 1$ /* adjust position */
\ENDWHILE
\WHILE{$left > 1$}
    \STATE $prev\_base \gets fasta.\texttt{fetch}(chrom, left-2, left-1)$ /* get preceding base */
    \IF{deletion and $seq[-1] == prev\_base$}
        \STATE $seq \gets prev\_base + seq[:-1]$ /* shift deletion left */
        \STATE $left \gets left - 1$ /* update position */
    \ELSIF{insertion and $seq\_alt[-1] == prev\_base$}
        \STATE $seq\_alt \gets prev\_base + seq\_alt[:-1]$ /* shift insertion left */
        \STATE $left \gets left - 1$ /* update position */
    \ELSE
        \STATE \textbf{break} /* cannot shift further */
    \ENDIF
\ENDWHILE
\STATE \textbf{Return:} $(left, seq, seq\_alt)$
\end{algorithmic}
\end{algorithm}

\section{Pileup image generation}

The \texttt{PileupFeaturizer} algorithm converts genomic variants into multi-channel images for CNN processing. It creates a 6-channel tensor where each channel captures different alignment properties: base identity (A/C/G/T encoded as intensity values), base quality scores, mapping quality, read strand direction, and matches to alternate/reference alleles. The algorithm centers a fixed-width window around each variant position, extracts aligned reads from the BAM file, sorts them by quality, and populates the image tensor row-by-row. The reference sequence occupies the bottom row with maximum quality values, while aligned reads fill remaining rows based on their CIGAR alignment coordinates. This transforms raw sequencing data into structured image format suitable for deep learning variant classification.

\begin{algorithm}
\caption{PileupFeaturizer}
\label{alg:pileupfeaturizer}
\begin{algorithmic}
\STATE \textbf{Input:} BAM file $B$, FASTA file $F$, candidates $C$
\STATE \textbf{Output:} Multi-channel pileup images dataset
\STATE $n \gets |C|$ /* number of candidate variants */
\STATE $X \gets \texttt{zeros}(n, channels, height, window)$ /* image tensor */
\STATE $y \gets \texttt{zeros}(n)$ if labeled else None /* labels if training */
\FOR{$i = 0$ to $n-1$}
    \STATE $chrom, pos, ref, alt \gets C[i][0:4]$ /* extract variant info */
    \STATE $start \gets pos - window/2$ /* define window boundaries */
    \STATE $end \gets pos + window/2 + 1$
    \STATE $ref\_seq \gets \texttt{fetch\_reference}(F, chrom, start, end)$ /* get reference */
    \STATE $reads \gets \texttt{fetch\_reads}(B, chrom, start, end)$ /* get aligned reads */
    \STATE $reads \gets \texttt{sort}(reads, \text{by mapping quality})$ /* prioritize high-quality reads */
    \STATE $pile \gets \texttt{zeros}(channels, height, window)$ /* initialize image */
    \FOR{$col = 0$ to $window-1$}
        \STATE $pile[0, height-1, col] \gets \texttt{base\_to\_intensity}(ref\_seq[col])$ /* reference base */
        \STATE $pile[1:5, height-1, col] \gets [1.0, 1.0, 1.0, \texttt{alt\_match}]$ /* reference row */
    \ENDFOR
    \FOR{$row = 0$ to $height-2$}
        \STATE $read \gets reads[row]$ /* process each read */
        \FOR{each aligned position $(qpos, rpos)$ in $read$}
            \IF{$start \leq rpos < end$}
                \STATE $col \gets rpos - start$ /* column in image */
                \STATE $base \gets read.sequence[qpos]$ /* read base */
                \STATE $pile[0, row, col] \gets \texttt{base\_to\_intensity}(base)$ /* base identity */
                \STATE $pile[1, row, col] \gets read.quality[qpos] / 40.0$ /* base quality */
                \STATE $pile[2, row, col] \gets read.mapping\_quality / 60.0$ /* mapping quality */
                \STATE $pile[3, row, col] \gets 0.0$ if reverse else $1.0$ /* strand */
                \STATE $pile[4, row, col] \gets 1.0$ if $base == alt$ else $0.0$ /* alt match */
                \STATE $pile[5, row, col] \gets 1.0$ if $base == ref$ else $0.0$ /* ref match */
            \ENDIF
        \ENDFOR
    \ENDFOR
    \STATE $X[i] \gets pile$ /* store completed image */
    \IF{labeled}
        \STATE $y[i] \gets C[i][-1]$ /* extract label if training */
    \ENDIF
\ENDFOR
\STATE \textbf{Return:} $\texttt{NumpyDataset}(X, y)$
\end{algorithmic}
\end{algorithm}

\section{DeepChem-Variant}
The DeepChem-Variant algorithm implements a complete variant calling pipeline using deep learning. It first extracts candidate variants from aligned reads using frequency thresholds, then generates multi-channel pileup images around each candidate site. These images are processed through a convolutional neural network (MobileNetV2) in batches to predict genotype probabilities (reference, heterozygous, or homozygous alternate). The algorithm computes genotype quality scores from prediction confidence, filters out reference calls, and writes remaining variants to a standard VCF file with proper formatting and metadata.

\begin{algorithm}
\caption{DeepChem-Variant}
\label{alg:deepchemvariant}
\begin{algorithmic}
\STATE \textbf{Input:} BAM file $B$, FASTA file $F$, output VCF path $O$
\STATE \textbf{Output:} VCF file with variant calls
\STATE $candidates \gets \texttt{CandidateFeaturizer}(B, F)$ /* extract potential variants */
\IF{$candidates$ is empty}
    \STATE $\texttt{write\_empty\_vcf}(O, F)$ /* create empty VCF with header */
    \STATE \textbf{Return:} $O$
\ENDIF
\STATE $pileup\_images \gets \texttt{PileupFeaturizer}(B, F, candidates)$ /* generate 6-channel images */
\STATE $predictions \gets []$ /* initialize prediction array */
\FOR{$i = 0$ to $|pileup\_images|$ step $batch\_size$}
    \STATE $batch \gets pileup\_images[i:i+batch\_size]$ /* create batch */
    \STATE $batch\_preds \gets \texttt{MobileNetV2}(batch)$ /* predict genotype probabilities */
    \STATE $predictions.\texttt{append}(batch\_preds)$ /* collect predictions */
\ENDFOR
\STATE $all\_predictions \gets \texttt{concatenate}(predictions)$ /* combine batches */
\STATE $genotypes \gets \texttt{argmax}(all\_predictions)$ /* most likely genotype */
\STATE $gq \gets \texttt{compute\_quality}(all\_predictions)$ /* genotype quality scores */
\STATE $\texttt{write\_vcf\_header}(O, F, sample\_name)$ /* write VCF header */
\FOR{each $(candidate_i, genotype_i, quality_i) \in \texttt{zip}(candidates, genotypes, gq)$}
    \IF{$genotype_i == 0$}
        \STATE \textbf{continue} /* skip reference calls */
    \ENDIF
    \STATE $chrom, pos, ref, alt \gets candidate_i[0:4]$ /* extract variant info */
    \STATE $gt\_string \gets$ "0/1" if $genotype_i == 1$ else "1/1" /* format genotype */
    \STATE $\texttt{write\_vcf\_record}(O, chrom, pos, ref, alt, gt\_string, quality_i)$ /* write variant */
\ENDFOR
\STATE \textbf{Return:} $O$
\end{algorithmic}
\end{algorithm}

\section{MobileNetV2}
The complete MobileNetV2 architecture implements efficient mobile computer vision. It begins with standard convolution for initial feature extraction, then processes through configurable inverted residual blocks that balance accuracy and computational efficiency. Each block configuration [t, c, n, s] specifies expansion ratio, output channels, repetition count, and stride. Width multipliers enable scaling model capacity for different resource constraints. Final layers include high-dimensional feature mapping (1280 channels), global average pooling for spatial dimension reduction, and linear classification head.

\begin{algorithm}
\caption{MobileNetV2 Network}
\label{alg:mobilenetv2}
\begin{algorithmic}
\STATE \textbf{Input:} Image $I$ with $C_{in}$ channels, input size $H \times W$
\STATE \textbf{Params:} Width multiplier $\alpha$, class count $N_{\text{class}}$
\STATE $x \gets \text{ConvBNReLU}(I, \text{out\_channels}=32, \text{stride}=2)$ /* initial feature extraction */
\FOR{each block configuration $[t, c, n, s]$ in settings}
    \STATE $C_{out} \gets \lceil \frac{c \times \alpha}{8} \rceil \times 8$ /* apply width multiplier */
    \FOR{$i = 1$ to $n$}
        \IF{$i = 1$}
            \STATE $x \gets \text{InvertedResidual}(x, C_{out}, \text{stride}=s, \text{expand\_ratio}=t)$ /* first block with stride */
        \ELSE
            \STATE $x \gets \text{InvertedResidual}(x, C_{out}, \text{stride}=1, \text{expand\_ratio}=t)$ /* subsequent blocks */
        \ENDIF
    \ENDFOR
\ENDFOR
\STATE $x \gets \text{ConvBNReLU}(x, \text{out\_channels}=1280)$ /* final feature mapping */
\STATE $x \gets \text{Mean}(x, \text{over spatial dims})$ /* global average pooling */
\STATE $y \gets \text{Linear}(x, \text{out\_features}=N_{\text{class}})$ /* classification layer */
\STATE \textbf{Return:} Class logits $y$
\end{algorithmic}
\end{algorithm}

\subsection{ConvBNReLU block}
The ConvBNReLU block is a standard convolutional building block optimized for mobile inference. The 3×3 convolution extracts spatial features while batch normalization stabilizes training and inference. ReLU6 activation (min(max(0, x), 6)) provides bounded non-linearity that improves quantization precision for mobile deployment, reducing numerical precision requirements compared to unbounded ReLU.

\begin{algorithm}
\caption{ConvBNReLU Block}
\label{alg:convbnrelu}
\begin{algorithmic}
\STATE \textbf{Input:} Feature map $x$, input channels $C_{in}$, output channels $C_{out}$, stride $s$
\STATE $y \gets \text{Conv2D}(x, \text{kernel}=3 \times 3, \text{stride}=s, \text{padding}=1, \text{bias}=False)$ /* convolution */
\STATE $y \gets \text{BatchNorm}(y)$ /* normalize activations */
\STATE $y \gets \text{ReLU6}(y)$ /* bounded activation function */
\STATE \textbf{Return:} $y$
\end{algorithmic}
\end{algorithm}

\subsection{Inverted Residual Block}

The Inverted Residual block is a core MobileNetV2 innovation addressing traditional depthwise convolution limitations. The "inverted" design expands narrow input channels to higher dimensions (expansion phase), applies efficient depthwise convolution for spatial feature extraction (filtering phase), then compresses back to narrow output channels (projection phase). This pattern maintains information flow in high-dimensional space while keeping input/output narrow for efficiency. Linear bottlenecks (no activation after final projection) preserve information flow. Residual connections enable gradient flow and feature reuse when input/output dimensions align, following ResNet principles adapted for mobile efficiency.
\begin{algorithm}
\caption{Inverted Residual Block}
\label{alg:invertedresidual}
\begin{algorithmic}
\STATE \textbf{Input:} Feature map $x$, input channels $C_{in}$, output channels $C_{out}$, stride $s$, expand ratio $t$
\STATE $C_{hidden} \gets C_{in} \times t$ /* calculate expanded channels */
\STATE $\text{use\_residual} \gets (s = 1) \land (C_{in} = C_{out})$ /* residual only if dimensions match */
\IF{$t = 1$}
    \STATE $y \gets \text{DepthwiseConv}(x, \text{kernel}=3 \times 3, \text{stride}=s)$ /* no expansion needed */
    \STATE $y \gets \text{BatchNorm}(y)$ /* normalize */
    \STATE $y \gets \text{ReLU6}(y)$ /* activate */
    \STATE $y \gets \text{PointwiseConv}(y, C_{out})$ /* project to output channels */
    \STATE $y \gets \text{BatchNorm}(y)$ /* normalize projection */
\ELSE
    \STATE $y \gets \text{PointwiseConv}(x, C_{hidden})$ /* expand channels */
    \STATE $y \gets \text{BatchNorm}(y)$ /* normalize expansion */
    \STATE $y \gets \text{ReLU6}(y)$ /* activate expanded features */
    \STATE $y \gets \text{DepthwiseConv}(y, \text{kernel}=3 \times 3, \text{stride}=s)$ /* spatial filtering */
    \STATE $y \gets \text{BatchNorm}(y)$ /* normalize filtering */
    \STATE $y \gets \text{ReLU6}(y)$ /* activate filtered features */
    \STATE $y \gets \text{PointwiseConv}(y, C_{out})$ /* project to output */
    \STATE $y \gets \text{BatchNorm}(y)$ /* normalize final projection */
\ENDIF
\IF{\text{use\_residual}}
    \STATE $y \gets x + y$ /* add residual connection */
\ENDIF
\STATE \textbf{Return:} $y$
\end{algorithmic}
\end{algorithm}

\section{InceptionV3}

InceptionV3 architecture implements the network-in-network design philosophy, where convolutional filters of various sizes (1×1, 3×3, 5×5) are applied in parallel within each module to capture features at multiple scales. The architecture systematically processes images through a hierarchical feature extraction pipeline that begins with stem convolutions for low-level feature extraction, progresses through InceptionA modules for multi-scale pattern recognition, utilizes reduction modules to compress spatial dimensions while expanding channel depth, employs InceptionC modules with factorized 7×7 convolutions for efficient mid-level feature processing, and concludes with InceptionE modules for high-level abstraction. The network incorporates factorized convolutions that break down larger convolutions into smaller, more efficient sequences (such as decomposing 3×3 into 1×3 and 3×1), significantly reducing computational complexity while maintaining representational power. Auxiliary classifiers are strategically placed to provide intermediate supervision during training, acting as regularizers that combat vanishing gradients in deep networks. Dimensionality reduction techniques are employed throughout to control computational complexity without sacrificing model expressiveness.

\begin{algorithm}
\caption{InceptionV3}
\label{alg:inceptionv3}
\begin{algorithmic}
\STATE \textbf{Input:} Image $x$ with $C_{in}$ channels, size $299 \times 299$
\STATE \textbf{Output:} Class logits and auxiliary logits (if training)
\STATE $x \gets \text{Conv2d\_1a\_3x3}(x)$ /* initial stem convolution */
\STATE $x \gets \text{Conv2d\_2a\_3x3}(x)$ /* stem progression */
\STATE $x \gets \text{Conv2d\_2b\_3x3}(x)$ /* stem completion */
\STATE $x \gets \text{MaxPool2d}(x, 3, 2)$ /* spatial downsampling */
\STATE $x \gets \text{Conv2d\_3b\_1x1}(x)$ /* channel reduction */
\STATE $x \gets \text{Conv2d\_4a\_3x3}(x)$ /* feature extraction */
\STATE $x \gets \text{MaxPool2d}(x, 3, 2)$ /* spatial downsampling */
\FOR{$i = 5b$ to $5d$}
    \STATE $x \gets \text{InceptionA}(x)$ /* parallel multi-scale convolutions */
\ENDFOR
\STATE $x \gets \text{InceptionB}(x)$ /* reduction with stride 2 */
\FOR{$i = 6b$ to $6e$}
    \STATE $x \gets \text{InceptionC}(x)$ /* factorized 7×7 convolutions */
\ENDFOR
\IF{training and aux\_logits}
    \STATE $aux \gets \text{InceptionAux}(x)$ /* auxiliary classifier */
\ENDIF
\STATE $x \gets \text{InceptionD}(x)$ /* reduction with stride 2 */
\FOR{$i = 7b$ to $7c$}
    \STATE $x \gets \text{InceptionE}(x)$ /* high-level feature extraction */
\ENDFOR
\STATE $x \gets \text{AdaptiveAvgPool2d}(x, (1, 1))$ /* global pooling */
\STATE $x \gets \text{Flatten}(x)$ /* vectorize features */
\STATE $x \gets \text{Dropout}(x)$ /* regularization */
\STATE $x \gets \text{Linear}(x, num\_classes)$ /* classification head */
\STATE \textbf{Return:} $x$ (and $aux$ if training)
\end{algorithmic}
\end{algorithm}

\subsection{BasicConv2d}
This fundamental building block combines three essential operations: convolution for spatial feature extraction, batch normalization for training stability and gradient flow optimization, and ReLU activation for introducing non-linearity while preserving gradient propagation. The bias-free convolution design leverages batch normalization's inherent bias handling, reducing parameter redundancy.
\begin{algorithm}
\caption{BasicConv2d}
\label{alg:basicconv2d}
\begin{algorithmic}
\STATE \textbf{Input:} Feature map $x$, output channels $C_{out}$, kernel params
\STATE $x \gets \text{Conv2d}(x, C_{out}, \text{bias=False})$ /* convolution without bias */
\STATE $x \gets \text{BatchNorm2d}(x)$ /* normalize activations */
\STATE $x \gets \text{ReLU}(x)$ /* non-linear activation */
\STATE \textbf{Return:} $x$
\end{algorithmic}
\end{algorithm}

\subsection{InceptionA}
This multi-scale feature extraction module implements parallel processing branches with different receptive fields to capture diverse spatial patterns. The 1×1 branch captures point-wise features and cross-channel correlations, the 5×5 branch (preceded by 1×1 reduction) captures medium-scale spatial patterns, the double 3×3 branch efficiently approximates larger receptive fields while reducing computational cost, and the pooling branch preserves existing feature representations. Feature concatenation combines these diverse representations into a unified output tensor.

\begin{algorithm}
\caption{InceptionA Module}
\label{alg:inceptiona}
\begin{algorithmic}
\STATE \textbf{Input:} Feature map $x$, pool features count
\STATE $branch_{1×1} \gets \text{BasicConv2d}(x, 64, 1×1)$ /* direct 1×1 path */
\STATE $branch_{5×5} \gets \text{BasicConv2d}(x, 48, 1×1)$ /* 5×5 reduction */
\STATE $branch_{5×5} \gets \text{BasicConv2d}(branch_{5×5}, 64, 5×5)$ /* 5×5 convolution */
\STATE $branch_{3×3} \gets \text{BasicConv2d}(x, 64, 1×1)$ /* double 3×3 reduction */
\STATE $branch_{3×3} \gets \text{BasicConv2d}(branch_{3×3}, 96, 3×3)$ /* first 3×3 */
\STATE $branch_{3×3} \gets \text{BasicConv2d}(branch_{3×3}, 96, 3×3)$ /* second 3×3 */
\STATE $branch_{pool} \gets \text{AvgPool2d}(x, 3, 1, 1)$ /* pooling path */
\STATE $branch_{pool} \gets \text{BasicConv2d}(branch_{pool}, pool\_features, 1×1)$ /* pool projection */
\STATE $output \gets \text{Concatenate}([branch_{1×1}, branch_{5×5}, branch_{3×3}, branch_{pool}])$ /* combine paths */
\STATE \textbf{Return:} $output$
\end{algorithmic}
\end{algorithm}

\subsection{InceptionB}
This architectural transition module reduces spatial resolution from 35×35 to 17×17 while expanding channel depth from 288 to 768. Multiple reduction paths maintain feature diversity during downsampling: direct 3×3 convolution with stride-2 for efficient reduction, double 3×3 path for complex pattern preservation, and max pooling for spatial downsampling. The increased channel count compensates for spatial information loss.
\begin{algorithm}
\caption{InceptionB Module}
\label{alg:inceptionb}
\begin{algorithmic}
\STATE \textbf{Input:} Feature map $x$
\STATE $branch_{3×3} \gets \text{BasicConv2d}(x, 384, 3×3, \text{stride=2})$ /* direct reduction */
\STATE $branch_{dbl} \gets \text{BasicConv2d}(x, 64, 1×1)$ /* double 3×3 path */
\STATE $branch_{dbl} \gets \text{BasicConv2d}(branch_{dbl}, 96, 3×3)$ /* expand channels */
\STATE $branch_{dbl} \gets \text{BasicConv2d}(branch_{dbl}, 96, 3×3, \text{stride=2})$ /* reduce spatial */
\STATE $branch_{pool} \gets \text{MaxPool2d}(x, 3, 2)$ /* pooling reduction */
\STATE $output \gets \text{Concatenate}([branch_{3×3}, branch_{dbl}, branch_{pool}])$ /* combine reductions */
\STATE \textbf{Return:} $output$
\end{algorithmic}
\end{algorithm}

\subsection{InceptionC}
This computational optimization module factorizes expensive 7×7 convolutions into more efficient asymmetric sequences. The 1×7 followed by 7×1 factorization reduces parameters from 49 to 14 while maintaining equivalent receptive field coverage. The double factorization path provides additional feature diversity through repeated asymmetric convolutions, enabling complex pattern recognition with reduced computational overhead.

\begin{algorithm}
\caption{InceptionC Module}
\label{alg:inceptionc}
\begin{algorithmic}
\STATE \textbf{Input:} Feature map $x$, 7×7 channel count
\STATE $branch_{1×1} \gets \text{BasicConv2d}(x, 192, 1×1)$ /* direct path */
\STATE $branch_{7×7} \gets \text{BasicConv2d}(x, channels_{7×7}, 1×1)$ /* 7×7 factorization */
\STATE $branch_{7×7} \gets \text{BasicConv2d}(branch_{7×7}, channels_{7×7}, 1×7)$ /* factorize to 1×7 */
\STATE $branch_{7×7} \gets \text{BasicConv2d}(branch_{7×7}, 192, 7×1)$ /* factorize to 7×1 */
\STATE $branch_{dbl} \gets \text{BasicConv2d}(x, channels_{7×7}, 1×1)$ /* double 7×7 path */
\STATE $branch_{dbl} \gets \text{BasicConv2d}(branch_{dbl}, channels_{7×7}, 7×1)$ /* first factorization */
\STATE $branch_{dbl} \gets \text{BasicConv2d}(branch_{dbl}, channels_{7×7}, 1×7)$ /* second factorization */
\STATE $branch_{dbl} \gets \text{BasicConv2d}(branch_{dbl}, channels_{7×7}, 7×1)$ /* third factorization */
\STATE $branch_{dbl} \gets \text{BasicConv2d}(branch_{dbl}, 192, 1×7)$ /* final factorization */
\STATE $branch_{pool} \gets \text{AvgPool2d}(x, 3, 1, 1)$ /* pooling path */
\STATE $branch_{pool} \gets \text{BasicConv2d}(branch_{pool}, 192, 1×1)$ /* pool projection */
\STATE $output \gets \text{Concatenate}([branch_{1×1}, branch_{7×7}, branch_{dbl}, branch_{pool}])$ /* combine paths */
\STATE \textbf{Return:} $output$
\end{algorithmic}
\end{algorithm}

\subsection{InceptionD}
This second reduction stage transitions from 17×17 to 8×8 spatial resolution while expanding channels from 768 to 1280. The module combines direct 3×3 reduction for efficiency with complex 7×7×3 factorized paths that maintain information richness through sequential asymmetric convolutions followed by spatial reduction. This design preserves feature diversity while preparing for final high-level processing.
\begin{algorithm}
\caption{InceptionD Module}
\label{alg:inceptiond}
\begin{algorithmic}
\STATE \textbf{Input:} Feature map $x$
\STATE $branch_{3×3} \gets \text{BasicConv2d}(x, 192, 1×1)$ /* 3×3 reduction path */
\STATE $branch_{3×3} \gets \text{BasicConv2d}(branch_{3×3}, 320, 3×3, \text{stride=2})$ /* spatial reduction */
\STATE $branch_{7×7×3} \gets \text{BasicConv2d}(x, 192, 1×1)$ /* 7×7×3 path */
\STATE $branch_{7×7×3} \gets \text{BasicConv2d}(branch_{7×7×3}, 192, 1×7)$ /* factorize 1×7 */
\STATE $branch_{7×7×3} \gets \text{BasicConv2d}(branch_{7×7×3}, 192, 7×1)$ /* factorize 7×1 */
\STATE $branch_{7×7×3} \gets \text{BasicConv2d}(branch_{7×7×3}, 192, 3×3, \text{stride=2})$ /* final reduction */
\STATE $branch_{pool} \gets \text{MaxPool2d}(x, 3, 2)$ /* pooling reduction */
\STATE $output \gets \text{Concatenate}([branch_{3×3}, branch_{7×7×3}, branch_{pool}])$ /* combine reductions */
\STATE \textbf{Return:} $output$
\end{algorithmic}
\end{algorithm}

\subsection{InceptionE}
The module splits 3×3 convolutions into separate 1×3 and 3×1 branches, effectively doubling feature diversity by capturing horizontal and vertical patterns independently. Complex double-branch paths maximize representational capacity through parallel processing of complementary feature patterns, providing rich feature representations for final classification decisions.

\begin{algorithm}
\caption{InceptionE Module}
\label{alg:inceptione}
\begin{algorithmic}
\STATE \textbf{Input:} Feature map $x$
\STATE $branch_{1×1} \gets \text{BasicConv2d}(x, 320, 1×1)$ /* direct path */
\STATE $branch_{3×3} \gets \text{BasicConv2d}(x, 384, 1×1)$ /* 3×3 split preparation */
\STATE $branch_{3×3a} \gets \text{BasicConv2d}(branch_{3×3}, 384, 1×3)$ /* horizontal split */
\STATE $branch_{3×3b} \gets \text{BasicConv2d}(branch_{3×3}, 384, 3×1)$ /* vertical split */
\STATE $branch_{3×3} \gets \text{Concatenate}([branch_{3×3a}, branch_{3×3b}])$ /* combine splits */
\STATE $branch_{dbl} \gets \text{BasicConv2d}(x, 448, 1×1)$ /* double 3×3 path */
\STATE $branch_{dbl} \gets \text{BasicConv2d}(branch_{dbl}, 384, 3×3)$ /* expand features */
\STATE $branch_{dbla} \gets \text{BasicConv2d}(branch_{dbl}, 384, 1×3)$ /* horizontal split */
\STATE $branch_{dblb} \gets \text{BasicConv2d}(branch_{dbl}, 384, 3×1)$ /* vertical split */
\STATE $branch_{dbl} \gets \text{Concatenate}([branch_{dbla}, branch_{dblb}])$ /* combine splits */
\STATE $branch_{pool} \gets \text{AvgPool2d}(x, 3, 1, 1)$ /* pooling path */
\STATE $branch_{pool} \gets \text{BasicConv2d}(branch_{pool}, 192, 1×1)$ /* pool projection */
\STATE $output \gets \text{Concatenate}([branch_{1×1}, branch_{3×3}, branch_{dbl}, branch_{pool}])$ /* combine all paths */
\STATE \textbf{Return:} $output$
\end{algorithmic}
\end{algorithm}

\subsection{InceptionAux}
This auxiliary classifier module addresses vanishing gradient problems in deep networks by providing intermediate supervision during training. Positioned at the network's midpoint, it processes intermediate features through spatial reduction, channel manipulation, and classification layers. The auxiliary loss signal improves gradient flow to earlier network layers, enhancing training convergence and preventing gradient degradation. During inference, this module is bypassed to maintain computational efficiency.
\begin{algorithm}
\caption{InceptionAux Module}
\label{alg:inceptionaux}
\begin{algorithmic}
\STATE \textbf{Input:} Feature map $x$, number of classes
\STATE $x \gets \text{AvgPool2d}(x, 5, 3)$ /* spatial reduction */
\STATE $x \gets \text{BasicConv2d}(x, 128, 1×1)$ /* channel reduction */
\STATE $x \gets \text{BasicConv2d}(x, 768, 5×5)$ /* feature expansion */
\STATE $x \gets \text{AdaptiveAvgPool2d}(x, (1, 1))$ /* global pooling */
\STATE $x \gets \text{Flatten}(x)$ /* vectorize */
\STATE $x \gets \text{Linear}(x, num\_classes)$ /* classify */
\STATE \textbf{Return:} $x$
\end{algorithmic}
\end{algorithm}

\end{document}